\documentstyle[11pt,fleqn]{article}
\def\titolo{\par\bigskip\begin{center}\bf\LARGE}
\def\endtitolo{\end{center}\par\bigskip\par\rm\normalsize}
\def\instit{\begin{center}\large}
\def\endinstit{\end{center}\rm\normalsize}
\def\references{\begin{thebibliography}{10}}
\def\endreferences{\end{thebibliography}\end{document}}

\newcommand{\btit}{\begin{titolo}}
\newcommand{\etit}{\end{titolo}}

\renewcommand{\author}[1]{\begin{center}\Large #1\end{center}}
\renewcommand{\date}[1]{\par\bigskip\par\sl\hfill #1\par\medskip\par}

\newcommand{\babs}{\hrule\par\begin{description}\item{Abstract: }\it}
\newcommand{\eabs}{\par\end{description}\hrule\par\medskip\rm}

\newcommand{\hs}{\qquad\qquad}         %%%  horizontal space
\newcommand{\nn}{\nonumber}            %%%  no number for eqnarray
\newcommand{\beq}{\begin{eqnarray}}    %%%  begequation/eqnarray
\newcommand{\eeq}{\end{eqnarray}}      %%%  endequation/eqnarray
\newcommand{\beqn}{\begin{eqnarray}}   %%%  begequation/eqnarray
\newcommand{\eeqn}{\end{eqnarray}}     %%%  endequation/eqnarray
\newcommand{\R}{\mbox{I$\!$R}}                   %%% real numbers
\newcommand{\ii}{\infty}                         %%% infinit
\newcommand{\Tr}{\,\mbox{Tr}\,}                  %%% Trace
\renewcommand{\Re}{\,\mbox{Re}\,}                %%% Real
\newcommand{\al}{\alpha}
\newcommand{\be}{\beta}

\newcommand{\de}{\delta}

\newcommand{\la}{\lambda}

\newcommand{\om}{\omega}

\newcommand{\th}{\theta}

\newcommand{\Ga}{\Gamma}

\newcommand{\La}{\Lambda}

\newcommand{\Om}{\Omega}

\begin{document}
%\tableofcontents       %%%%%%   index of section

\title{Thermal states in Anti-de Sitter space-time and the role of time-like
infinity}

\endtitle

\author{Luciano Vanzo \\}
\begin{instit}
Dipartimento di Fisica, Universit\'a di Trento, Italia\\
Istituto Nazionale di Fisica Nucleare, Gruppo collegato di trento
\end{instit}
\author{Giuseppe Turco \\}
\begin{instit}
Istituto Superiore di Studi Avanzati (SISSA)\\
Miramare, Trieste, Italia
\end{instit}
\vspace{1cm}

\babs
The grand-canonical thermodynamic potential for Bose and Fermi fields
in anti-de Sitter space-time is introduced and a Mellin complex
representation is given. This is used to investigate the
high temperature properties of the thermal state at finite charge.
The time-like nature of spatial infinity is shown to act as a boundary,
and changes the free energy of the field in thermal equilibrium.
\eabs
\vspace{1cm}

Quantum field theory in anti-de Sitter space-time (AdS)
has been discussed over the years by several authors
with several motivations
\cite{fron65-37-221,avis78-18-3565,brei82-144-249,burg85-153-137,filt91-43-485}.
The lack of global hyperbolicity and its effects
on quantum fields was the main source of curiosity in the early discussion
of AdS \cite{avis78-18-3565}. There it was shown that an acceptable
quantum theory can be
constructed which has positive energy. Shortly after these investigations,
it was found that AdS space-time occurs as the ground state geometry in
certain supergravity theories with gauged $SO(N)$ internal symmetry,
$N\geq 4$ \cite{brei82-144-249,nico81-108-285,nico81-188-98}.
In these theories the ground state was shown to be stable. Stability was also
established for gravity
fluctuations about the AdS background \cite{abbo82-195-76}.
One can observe, at this point, that AdS
space-time is not so unfrequent as it looks at first glance. In fact,
any theory containing a scalar-gravity sector with action such as

\beq
S=-\frac{1}{16\pi G}\int\sqrt{-g}Rd^4x+\int d^4x\sqrt{-g}\left\{\frac{1}{2}
g^{\mu \nu}\partial_{\mu}\phi\partial_{\nu}\phi -V(\phi)\right\}
\label{act}
\eeq
will have AdS background geometry with constant $\phi =\phi_0$, if
$\phi_0$ is a critical point of $V(\phi)$ such that $V(\phi_0)<0$.
This could be the case for a potential with tree-level, spontaneous
symmetry breaking. The tree level cosmological constant is then
$\Lambda =-8\pi GV(\phi_0)$ and the scalar curvature $R=4\La$.
A more complicated model is the
self-interacting scalar theory defined by the classical action

\beq
S=-\frac{1}{16\pi G}\int\sqrt{-g}Rd^4x+\int d^4x\sqrt{-g}
\left\{\frac{1}{2}g^{ab}\partial_a\psi \partial_b\psi -\frac{1}{2}
(m^2-\xi R)\psi^2-\frac{\la}{4!}\psi^4\right\}
\label{act1}
\eeq
Looking for a critical point with anti-de Sitter background
solution and constant $\psi =\psi_0$ we found the following relations

\beq
\psi_{0}^{2}=-\frac{6m^2}{48\pi Gm^2\xi +\la} \hs R=\frac{48\pi Gm^4}{48\pi
Gm^2\xi+\la}
\label{psr}
\eeq
Then either $m^2 >0$ and we have a de Sitter solution or $m^2<0$ and we
have anti-de Sitter solution with tree-level spontaneous symmetry breaking.
Notice that the cosmological constant is not given
by the minimum of the potential if $\xi\neq 0$. Finally, if $m^2=0$ then the
only maximally symmetric solution is Minkowski space-time. The one-loop
effective potential of the $\la\phi^4$ interaction has been computed in the
AdS background \cite{burg85-153-137} and
was found real at the point of inflection of the classical potential.
Another sign of this remarkable
stability property of AdS gravity, we have not found any bounce solution of
the Euclidean field equation which interpolates the two classical vacua.
We mention, at this point, that the
renormalization of $\phi^4$ interaction carries over to AdS space-time in
a simple way \cite{filt91-43-485}, and the same is expected for gauge
theories, notably QED and QCD. The one-loop effective potential for higher
spin fields have also been computed in the Euclidean formalism
\cite{camp91-43-3958,camp92-45-3591}, in 2-dimension
\cite{saka85-255-401} and arbitrary dimension \cite{inam85-73-1051} in
connection with possible supersymmetry breaking.
At last, anti-de Sitter and Weyl conformal gravity have been recently proved
to be the only type of gravity which have a consistent interaction with
massless higher spin fields \cite{frad87-177-63}. It was then shown that
consistent
AdS strings exist for any $D\neq 26$ (or $D\neq 10$) provided the cosmological
constant has that critical value which is required by anomaly
cancellation \cite{frad91-261-26}.
So, although AdS space-time do not seems to correspond to our present world,
its unusual properties and occurence in spontaneous symmetry breaking
theories makes it more than a curiosity.
The deeper question as to whether an AdS phase was possible in the
early Universe seems to be a very interesting problem.
Such a phase could be possible after the electroweak phase transition,
since as the
Universe cools down a small positive cosmological constant could develop
dynamically from the Higgs field. After all, the present upper
bound $\mid\Lambda\mid\leq 10^{-54} cm^{-2}$, leaves enough room for a
non zero $\Lambda$.
This was proved to occur in a simpler
model of vacuum decay in Ref.\cite{cole80-21-3305}, although the AdS regions
of true vacuum were shown to be dynamically unstables.
The possibility of an AdS phase
would have several interesting effects on the Universe history. Among
these, we mention the possibility that left-handed/right-handed neutrinos
are distinguished by a superselection rule and the known
difficulties to accomodate monopoles in AdS \cite{fron75-12-3819}.

Our curvature conventions are as in Hawking-Ellis \cite{hawk73b}, but the
metric signature which is $(+,-,-,-)$. We use also $\hbar=c=1$ so
that $G^{-1}=m_{p}^{2}$ is the Planck mass.
Anti-de Sitter space is a maximally symmetric and globally static
solution of vacuum Einstein equations with positive cosmological constant
(this is merely a matter of choice)
\beq
R_{ab}-\frac{1}{2}Rg_{ab}+\La g_{ab}=0
\label{ein}
\eeq
The condition is $R_{abcd}=k[g_{ac}g_{bd}-g_{ad}g_{bc}]$ and then from
Einstein
equations $3k=\La$. One convenient form of the anti-de Sitter metric
tensor is
\beq
ds^2=a^{-2}(\cos\rho)^{-2} \{dt^2-[d\rho^2 +\sin^2\rho (d\th^2+\sin^2\th
d\phi^2)]\}
\label{met}
\eeq
where $a$ is a parameter that sets the scale of geometry,
$0\leq\rho <\pi/2$, $\th$ and $ \phi$ are coordinates on the
2-sphere and
$t\in(-\infty,\infty)$. The space-time which is
obtained identifying $t$ with $t+n\pi$, $n$ integer, is the commonly known
anti-de Sitter space-time, and clearly contains closed time-like geodesics.
The symmetry group is $SO(3,2)$ (or rather its
covering) and the unitary,
infinite dimensional representations are well known
after the systematic treatment given by Fronsdal \cite{fron65-37-221}, Evans
\cite{evan67-8-170} and earlier by Ehrman \cite{ehrm56-53-290}.
Many properties can be read of Eq.~(\ref{met}). Under the Wick
rotation $t\rightarrow -i\tau$, the metric
goes over to the line element of hyperbolic four space.
The metric is also conformal
to half the Einstein static Universe (EsU), namely the
cylinder $\R\times S^3$ with the product Lorentz metric.
The equator of the EsU three sphere, that is the set $\rho=\pi/2$, correspond
to spatial infinity in
AdS. The normal vector field to the surface $\rho=\pi/2$ is
space-like, i.e. spatial infinity is time-like in AdS, much like the boundary
of any finite spatial region in Minkowski space-time. It is this feature that
makes unusual the causal structure of AdS, which is not globally hyperbolic
\cite{hawk73b}. Finally, the scalar curvature is $R=12a^2=4\Lambda$.

The quantum theory of a complex scalar field coupled to AdS gravity as
in Eq.~(\ref{act1}) with $\la=0$, can be considered well understood. Since
the space-time fails to possess any global Cauchy surface, the initial value
problem is well posed if suitable boundary conditions are imposed at
infinity, or equivalently on the equator of the EsU three-sphere. These
boundary conditions can be determined from the physical requirement that the
formally conserved Klein-Gordon scalar product and the Killing energy
are actually conserved for any mode function
\cite{avis78-18-3565,brei82-144-249}.
With the help of the parameters $\eta= (m^2-\xi R)a^{-2}$ and
$\be_{\pm}=[3\pm (9+4\eta)^{1/2}]/2$, the one-particle energies are given by
\beq
\om^{\pm}_{k}=a\be_{\pm}+ak \hs k=0,1,2,...,\ii
\label{spe}
\eeq
and each $\om^{\pm}_{k}$ has degeneration $d(k)=(k+1)(k+2)/2$. The
corresponding modes, say $\psi_{J}^{\pm}$, form a basis for the two
unitary, spin zero
representations $D(\be_{\pm},0)$ of $SO(3,2)$ described by Fronsdal
\cite{fron75-12-3819}.
Reality of the frequencies gives now the restriction $\eta >-9/4$.
The convergence of the Killing energy
can be obtained with two different conditions: (i)
either $2\xi=\be_-/(2\be_-+1)$ and $\be_->1/2$ and both kind of modes are
admissible or (ii) $\xi$ is arbitrary and only the $(+)$-modes are admissible.
If energy is defined using the canonical stress tensor, then its
conservation requires the further condition $2\be_{\pm}>3$.
Then only the $(+)$-modes are admissible, since $\be_-$ is never greater than
$3/2$.
In the case of conformal coupling, $m^2=0, \xi=1/6$, both kind of modes are
admissible and we have two inequivalent quantum theories. In this case
$\be_+=2$ and $\be_-=1$, but neither $D(2,0)$ nor
$D(1,0)$ carry any unitary representation of the conformal group for the
AdS metric (this group is $SO(2,4)$ as for flat Minkowski space).
Hence the conformal symmetry of the classical system is spontaneously
broken in the quantum theory \cite{fron75-12-3819}.
The Fock space will carry a
reducible unitary representation of $SO(3,2)$ with an invariant vacuum.
The quantization of a spinor field is performed along similar lines
\cite{fron75-12-3810,brei82-144-249}. The
Dirac equation can be solved by separation of variables and the
positive frequency spectrum is
\beq
\om_k=m+\frac{3a}{2}+ak \hs k=0,1,2...,\ii
\label{spe1}
\eeq
Each $\om_k$ has degeneration $d(k)=(k+1)(k+2)$, namely twice
the scalar degeneration. We leave aside the discussion of parity
assignments and many other interesting properties of spinors in AdS, as
the only thing of importance to us is just the frequency spectrum and its
degeneration.

If $J$ denotes a complete set of quantum numbers needed to specify a
one-particle state, the quantum hamiltonian for the Bose field is
\beq
H_b=\frac{1}{2}\sum_J\om^{\pm}_{J}[a_Ja^{\dag}_{J}+a^{\dag}_{J}a_J+
b_Jb^{\dag}_{J}+b^{\dag}_Jb_J]
\label{ham}
\eeq
After normal ordering the divergente vacuum energy
\beq
E^{b}_{vac}=\sum_{J}\om_{J} =\frac{1}{2}\sum_{k=0}^{\infty}(k+1)(k+2)
\om^{\pm}_{k}
\label{vae}
\eeq
will appear. The normal ordered charge operator is similarly given by
\beq
:Q_b:=\sum_{J}[a^{\dag}_{J}a_J-b^{\dag}_{J}b_J]
\label{nch}
\eeq
The energy functional for fermions gives the hamiltonian
\beq
H_f=\sum_{J}\om_J[a^{\dag}_{J}a_J-b_Jb^{\dag}_J]
\label{ham1}
\eeq
and the charge operator is the same as Eq.~(\ref{nch}).
Normal ordering of $H_f$ also gives the vacuum energy for fermions
\beq
E_{vac}^{f}=-\sum_{J}\om_{J}=-2\sum_{k=0}^{\infty}(k+1)(k+2)\om_{k}
\label{i1}
\eeq

Now a thermal equilibrium state at finite charge is defined
to be the density operator
\beq
U=Z^{-1}\exp[-\be(:\!H\!:-\mu:\!Q\!:)]
\label{den}
\eeq
where $Z=\Tr\exp[-\be(:\!H\!:-\mu:\!Q\!:)]$ is the partition function and $\mu$
the chemical potential. Also, the hamiltonian was normal ordered with respect
to
the Fock vacuum.
On spaces with infinite volume this usually cannot be realized as a density
matrix on the Hilbert-Fock space. The situation in AdS will prove to
be different. The thermodynamic potential is defined as
\beq
\Om(\be,\mu,a)=-\frac{1}{\be}\ln Z
\label{the}
\eeq
{}From Eqs.~(\ref{den}),(\ref{the}) and
the statistical definition of entropy it follows that
\beq
S=-\left(\frac{\partial\Om}{\partial T}\right)_{\mu,a}
\label{ent}
\eeq
and the expectation values of energy and charge are, respectively
\beq
E=<H>=TS+\mu Q+\Om \hs Q=-\left(\frac{\partial\Om}{\partial\mu}\right)_{\be,a}
\label{ene}
\eeq
The system also depends on the external parameter $a$, the inverse radius
of the Universe. The work done on the system to change from $a$ to $da$
at fixed temperature and charge is
\beq
\de W=\left(\frac{\partial\Om}{\partial a}\right)_{\be,\mu}da=-\frac{1}{3}a
\left(\frac{\partial\Om}{\partial a}\right)_{\be,\mu}\frac{dV}{V}
\label{wor}
\eeq
where we introduced half the volume of the Einstein static Universe,
namely $V=\pi^2a^{-3}$. That $V$ is the effective volume of the system
will be clear from the asymptotic properties of the thermal state both
at high $T$ as well as at small $a$ (small curvature). Thus we have a
pressure such that
\beq
PV=-\frac{1}{3}a\left(\frac{\partial\Om}{\partial a}\right)_{\be,\mu}
\leq \frac{E}{3}
\label{pre}
\eeq
The given inequality comes from the definitions and is saturated in
the massless limit. Hence the usual thermodynamics
identities are recovered but $\Om=-PV$. Rather, we should obtain a virial-like
expansion for $P$ as a function of $V$. This is because the thermal gas
is not an homogeneous system in the thermodynamical sense though space-time
is homogeneous in the geometric sense. In fact the gravitational
forces acting on the particles depend on position.

The next step would be to compute the partition function.
The AdS metric describes a static gravitational field without
horizons, and we observed that the appropriate Euclidean manifold is the
hyperbolic 4-space $H^4$ \cite{dowk76-13-224,camp91-43-3958}.
By quantizing
the Euclidean field theory on $H^4$ we founded that the Wick rotation of
the theory is unitarily equivalent to the quantum theory in AdS carrying
the $D(\be_+,0)$ representation of $SO(3,2)$ (see also
Ref.\cite{camp91-43-3958}). For some reason, the $D(\be_-,0)$
representation does not
admit an Euclidean formulation, at least on $H^4$. Moreover, the high
temperature expansion for the free energy was found harder in the
Euclidean formalism, where the frequency spectrum is continuous.
Hence we shall discuss the theory in space-time directly.
Using the known form of the spectrum and expanding the traces we get
\beq
\Om_{\pm}(\be,\mu,a)=\mp\frac{g_{\pm}}{\be}\sum_{k=0}^{\infty}
\frac{1}{2}(k+1)(k+2)
\ln \left(1\pm e^{-\be(\om_{k}+\mu)}\right)+[\mu \rightarrow -\mu]
\label{ome}
\eeq
where the subscripts $\pm $ refer to fermion/boson degrees of freedom
and $g_{\pm}$ is the spin degeneration factor, namely $g_-=1$, $g_+=2$.
The absolute convergence of the series implies that the density matrices are
trace-class even if the spatial volume is infinite. This is true because we
took
the hamiltonians in the normal ordered form by discarding the vacuum
energy. There are no compelling reasons to do so, other than to save
the $SO(3,2)$ invariance of the vacuum which seems to be spoiled by a finite
vacuum energy (see below).
The expected total charge in the ensemble is
\beq
Q_{\pm}(\be,z,a)=g_{\pm}\sum_{k=0}^{\infty}\frac{1}{2}(k+1)(k+2)
\left[\frac{1}{ze^{\be \om_k}\pm 1}-\frac{z}{e^{\be \om_k}\pm z}\right]
\label{cha}
\eeq
where $z=\exp(-\be\mu)$ is the fugacity. It is finite for the formal reason
that AdS is conformal to half EsU, whose spatial section is compact.
Alternatively, we can say it is a gravity effect due to time-like infinity.

There are several consequences of these facts. First, it is
unnecessary to perform the thermodinamic limit, since the system behaves
effectively as a finite volume system. Then the phenomenon of
Bose-Einstein condensation is not a phase transition
in a strict mathematical sense, namely, the first derivative of the specific
heat is continuous everywhere.
The macroscopic occupation of the ground state
is nevertheless present above a critical charge density and in that sense the
condensation phenomenon occurs. To justify our claims, let us suppose
$Q_->0$ (more anti-particles than particles). Then $\mu >0$ and so $z_0<z<1$,
where
$z_0=\exp(-\be \om_0)$ is the lower extreme value for $z$ consistent with
positive occupation numbers (the other extreme value is $z_1=\exp\be\om_0$).
Let us further define $Q_{e}=Q_--Q_0$, where $Q_0$ is the ground state charge
(the $k=0$-term in Eq.~(\ref{cha})). Clearly $Q_{e}$ is a monotonically
decreasing
function of $z$ at fixed $\be$. Moreover, since $\om_1>\om_0$ strictly,
$Q_{e}$ remains bounded as $z\rightarrow z_0$ (i.e. $\mu \rightarrow \om_0$)
, in contrast to $Q_-$ which diverges: $Q_{e}(\be , z)\leq Q_{e}(\be , z_0)$.
So when $Q_-$ exceedes the bound of $Q_{e}$ (almost) all the new particles will
go into the ground state, i.e. we do have condensation but the process is
mathematically smooth. In fact since the series in Eq.~(\ref{cha}), is
absolutely convergente with all of its derivatives we can
differentiate it with respect to $\be$
and $\mu$ any number of times in their permitted ranges, $0<\be <\infty$,
$-\om_0 <\mu <\om_0$, without encountering any singularity.
The critical total charge is $Q_c=Q_e(\be, z_0)$. Note that since everything
converge, the expectation value of the quantum field vanishes at all
temperatures.

We now compute $Q_{\pm}$
in the high temperature, relativistic regime, $a\be\ll 1$.
To look at this expansion, notice that the infinite series reprentations
of $Q_{\pm}(\be,z)$ and $\Om_{\pm}(\be,\mu)$ are almost useless since
the factors
$\be\om_k$ grow no matter how small $\be$ is. Hence we use a different
procedure, familiar to mathematicians, and which we think is interesting in
its own rights. Using the following Mellin transform pair into
Eq.~(\ref{cha})
\beq
\frac{1}{e^x\pm 1}=\frac{1}{2\pi i}\int_{c-i\ii}^{c+i\ii}\Ga (s)
\zeta_{\pm}(s)x^{-s}ds
\hs c>1
\eeq
where $\zeta_+(s)=(1-2^{1-s})\zeta(s)$ and $\zeta_-(s)=\zeta(s)$
is the Riemann zeta-function, we can
do the summations provided we shift the integration
contour to $\Re s=c>3$. This we can do because the integrand is analytic
in that region. Then we get the complex representations for $Q_{\pm}$
\beq
Q_{\pm}=\frac{g_{\pm}}{4\pi i}\int_{c-i\ii}^{c+i\ii}\Ga (s)\zeta_{\pm}(s)
(\be a)^{-s}\Xi(s,v_{\pm} )ds-[\mu \rightarrow -\mu] \hs c>3
\label{com}
\eeq
where we defined
\beq
\Xi(s,v_{\pm})=\zeta (s-2,v_{\pm})+(3-2v_{\pm})\zeta (s-1,v_{\pm})+
(v_{\pm}^{2}-3v_{\pm}+2)\zeta (s,v_{\pm})
\label{fun}
\eeq
Here $\zeta (s,v)$ denotes the Riemann-Hurwitz zeta-function
and $v_{\pm}=(\om^{\pm}_0-\mu)/a$, where $\om^{\pm}_0$ is the lowest
eigenvalue of particle spectrum (see Eqs.~(\ref{spe}),(\ref{spe1})).
The function $\zeta(s,v)$ is defined for $\Re s>1$ by
\beq
\zeta(s,v)=\sum_{k=0}^{\ii}(k+v)^{-s}
\label{zet}
\eeq
and by analytic continuation elsewhere. The analytically continued
function then has a simple pole
at $s=1$ of unit residue \cite{grad80b}. It follows that
$\Xi(s,v_{\pm})$ is a
meromorphic functions with simple poles at $s=1$, $s=2$ and $s=3$.
Moreover
\beq
\zeta(-n,v)=-\frac{B_{n+1}(v)}{n+1} \hs \zeta^{'}(0,v)=\ln\Ga(v)-
\frac{1}{2}\ln 2\pi
\label{pro}
\eeq
where $n=0,1,2,...,\ii$ and $B_n(v)$ are the Bernoulli polynomials
\cite{grad80b}.

A similar Mellin representation can also be given for $\Om_{\pm}(\be,\mu,a)$
in the form
\beq
\Om_{\pm}(\be,\mu,a)=-\frac{ag_{\pm}}{4\pi i}\int_{c-i\ii}^{c+i\ii}\Ga (s-1)
\zeta_{\pm}(s)(\be a)^{-s}\Xi(s-1,v_{\pm})ds+[\mu \rightarrow -\mu]
\label{comp}
\eeq
where now $c>4$. For any ultrastatic manifold, such a complex representation
of the free energy has been given in general using appropriate
zeta-functions \cite{alle86-33-3640,byts92-291-26}.
Now closing the contour to the left and evaluating the integral with the
method of residues we quickly get an expansion in inverse powers
of $\be =T^{-1}$, i.e. the
high temperature expansion we sought for. However, in this procedure we
clearly neglet the contribution of the integral along the circle at infinity.
As a consequence the high temperature expansions given below are only
asymptotically close to the true free energy, and in general they do not
converge. The first few terms for $\Om_{\pm}(\be,\mu,a)$ are given by
\begin{eqnarray}
\Om_-(\be,\mu,a)&=&-\frac{V\pi^2}{45}T^4-\frac{(3-2\be_+)\zeta(3)}{a^2}
T^3-\frac{V}{6}(m^2-\xi R+\mu^2+2a^2)T^2 \nonumber \\
& - & \frac{a}{2}[\Xi(0,v_+)+\Xi(0,v_-)]T\ln\left(\frac{T}{a}\right)-
\frac{a}{2}\left[\Xi^{'}(0,v_+)+\Xi^{'}(0,v_-)\right]\frac{T}{a} \nn \\
&-&\frac{a}{4}\left[\Xi(-1,v_+)+\Xi(-1,v_-)\right]+O\left(\frac{a}{T}\right)
\label{hit}
\end{eqnarray}
\begin{eqnarray}
\Om_+(\be,\mu,a)&=&-\frac{7V\pi^2}{180}T^4+\frac{3V\zeta(3)m}{2\pi^2}T^3
-\frac{V}{6}\left(m^2+\mu^2-\frac{a^2}{4}\right)T^2 \nonumber \\
& - & \ln 2[\Xi(0,v_+)+\Xi(0,v_-)]T
+\frac{a}{2}\left[\Xi(-1,v_+)+\Xi(-1,v_-)\right]+O\left(\frac{a}{T}\right)
\label{hit1}
\end{eqnarray}
The charge in the ensemble can be found from Eqs.~(\ref{hit}),(\ref{hit1})
using $Q_{\pm}=-\partial\Om_{\pm}/\partial\mu$. Then we get for example
\begin{eqnarray}
Q_-(\be ,\mu, a)&=&\frac{\mu V}{3}T^2+
\frac{(3-2\be_+)\mu}{a^2}T\ln\left(\frac{T}{a}\right)+\frac{(7-6\be_+)\mu}
{2a}\frac{T}{a}\\
&+&\frac{1}{2}\left[\left(\e+2+\frac{\mu^2}{a^2}+(3-2\be_+)\frac{\mu}{a}\right)
\Psi(\be_+-\mu/a)-[\mu\rightarrow -\mu]\right]\frac{T}{a} \nn \\
&-&\frac{1}{4}\left[\Xi(0,v_+)-\Xi(0,v_-)\right]+O\left(\frac{a}{T}\right)
\label{htc}
\end{eqnarray}
where $\Psi(x)$ is the logarithmic derivative of $\Ga(x)$. Following
Eq.~(\ref{pro}), the values $\Xi(0,v_{\pm})$ and $\Xi(-1,v_{\pm})$ are
computable polynomials of $\mu$, while for $\Xi^{'}(0,v_{\pm})$ a closed
formula is not available to the authors knowledge.
The $\ln(T/a)$ term is the contribution of the double pole of the
integrand at $s=1$ in Eq.~(\ref{comp}).
Once more, $V=\pi^2a^{-3}$ is half the volume of the spatial section
of EsU. Expressing $a$ through $V$, the series (\ref{hit}), (\ref{hit1})
then take the form of a virial-like expansion.
The fact that the effective volume of the theory is $V$, is in accord
with the general theory of the free energy on a static manifold
\cite{dowk78-11-895,dowk89-327-267,dowk89-39-1235,kirs91-8-2239}. By means
of a conformal
transformation to the ultrastatic, fictitious space-time whose metric tensor
is $h_{ab}=g_{ab}(g_{00})^{-1}$, the so called optical space-time
\cite{gibb78-358-467}, it
is seen that the expansion only depends on the metric properties of this last
manifold. In our case indeed, the optical space-time is the Einstein static
Universe, although the expansion of the free energy is very different in
that manifold \cite{park91-44-2421}.

To leading order, the critical charge density $Q_c$ and temperature $T_c$,
take the well known value \cite{bern91-66-683,toms92-69-1152,habe81-46-1497}
\beq
Q_c=\frac{V\om_0}{3}T^2 \hs T_c=\left(\frac{3Q}{V\om_0}\right)^{1/2}
\label{cri}
\eeq
where $\om_0=a\be_+$ is the lowest eigenvalue in the particle spectrum.
Notice that $T_c$ is finite in the massless limit $m^2=\xi=0$, as well as
in the conformally invariant case, $m^2=0$, $\xi=1/6$.

The $T^3$ and the $T\ln(T/a)$ terms are the new feature of the thermal
state: they should be absent on a
manifold whose boundary is empty but should be present
for fields confined in a cavity, for example. The factors multiplying these
terms are seen to be the Seeley-Schwinger-DeWitt asymptotic
coefficients of the conformally related wave operator in the Einstein
static Universe with Dirichlet or Neumann boundary conditions on the
equator of the EsU three sphere.
In this
sense we can say that $\Om_{\pm}(\be,\mu,a)$ receive contributions from
time-like infinity. The $T^2$ term, on the other hand, is
positive in flat space as well as in many other space-times, for example
in the Einstein static Universe \cite{park91-44-2421}. In AdS it can be
negative and the limit $a\rightarrow 0$ do not reproduces the flat space
results.

Let us finally compute the vacuum energy, also calculated for
$O(N)$-supergravity in the Ads background in Ref.\cite{alle83-124-353}.
It is different from zero at $N=4$ but it seems not to break supersymmetry
anyway \cite{saka84-146-38}.
The function $2^{-1}\Xi (s, \be_+)a^{-s}$ is just one possible regularized
expression for the boson vacuum energy since for $\Re s>3$
\beq
2^{-1}\Xi(s,\be_+)a^{-s}=\frac{1}{2}\sum_{k=0}^{\infty}(k+1)(k+2)\om_{k}^{-s}
\label{p}
\eeq
This has an analytic continuation which is regular at $s=-1$.
Thus $2^{-1}a\Xi (-1, \be_+)$ is well defined,
is the zeta-function definition of $E^{-}_{vac}$ and is not afflicted by any
anomalous scale dependence. In this way we obtain
\beq
E^{-}_{vac}=-\frac{a}{2}\left[\frac{(m^2a^{-2}-12\xi )^2}{12}+
\frac{m^2a^{-2}-12\xi}{4}+\frac{19}{120}\right]
\label{ben}
\eeq
Since $\eta =m^2a^{-2}-12\xi >-9/4$ we have $E^{-}_{vac}<0$ but in the interval
$-2.1<\eta <-0.9$, where it is positive. For fermions we get in the same way
\beq
E_{vac}^{+}=a\left[\frac{m^4}{6a^4}-\frac{m^2}{4a^2}+\frac{17}{480}\right]
\label{fen}
\eeq
as the statistic accounts for the opposite sign.
However, any $SO(3,2)$-invariant vacuum stress tensor will have
infinite total energy, being of the form $T_{ab}=\rho g_{ab}$, for some
constant $\rho$. Hence a finite vacuum energy seems to break the $SO(3,2)$
symmetry of the Fock vacuum. In fact, the ten $SO(3,2)$
generators, $L_{\mu\nu}$ and $P_{\mu}$, have the commutation relations of the
Poincar\'e group, only that
\beq
\left[P_{\mu},P_{\nu}\right]=-iaL_{\mu\nu} \hs [L_{\mu\nu},P_{\al}]=
ig_{\mu\al}P_{\nu}-ig_{\nu\al}P_{\mu}
\label{com}
\eeq
where $g_{\mu\nu}=$ diag$(+,-,-,-,+)$.
For $i,j=1,2,3$, we have $\left[L_{0i},P_j\right]=ig_{ij}H$, where
$H=P_0$ is the Hamiltonian.
The vacuum must then have zero energy for the symmetry to be unbroken, or what
is the same, the hamiltonian must be normal ordered to fulfill the commutation
relations. Still another way to see this is to note that since the vacuum
energy is a $c$-number, it can be made compatible with the symmetry if a
suitable central extension of the $SO(3,2)$ algebra can be found.
No such extension exists since the algebra is semi-simple.

In conclusion,
thermal states at finite charge in anti-de Sitter space-time certainly
exists in a well defined mathematical sense, even for a quantum field
theory. The time-like nature of spatial infinity strongly affects the
properties of the free energy. Among these, the high temperature expansion
is modified by boundary-like terms and the naive flat space limit do not
reproduces the flat space results. The Bose-Einstein phenomenon occurs but
it is not characterized by a well defined critical charge or temperature.
In particular, the expectation value of the quantum field in the thermal
state vanishes at all temperatures. These properties of course will modify
the usual picture of finite temperature symmetry braking (or restoration),
but to be more precise the effective potential at
finite temperature for an interacting theory should be computed.

\end{document}